\documentstyle[12pt,psfig]{article}
\begin{document}
%\draft
%\tightenlines
%\begin{large}
\title{Comparative study of spanning cluster distributions in different dimensions}
\author{Parongama Sen\\
%\address Department of Physics, Surendranath College, 24/2 Mahatma Gandhi Road,
 Department of Physics, Surendranath College,\\ 
24/2 Mahatma Gandhi Road, 
Calcutta 700009, India}
\maketitle
\begin{abstract}
The probability distributions of  the  masses of the clusters  
spanning from top to bottom 
%are obtained for
of a percolating lattice at the percolation  threshold are
obtained in all dimensions from  two to five.
The first two cumulants and the exponents for the
universal scaling  functions are shown to have simple power
law variations with the dimensionality.
The cases where multiple spanning clusters occur are discussed 
separately and compared.  
\end{abstract}
 
Percolation is a subject which has been studied extensively 
for the last few decades. The relevance of percolation in various areas
of physics is also well established.
Although many of the properties of percolating systems are well understood
and studied, there still remains a lot of details to be
explored and intricate questions to be 
addressed\cite{SA,DS1}. 

At the critical point (percolation threshold) there appears for the
first time a  cluster spanning the whole lattice. 
The spanning cluster is a fractal in the sense its mass $M$ scales
with the length as $L^D$ where $D < d$, $d$ being the spatial dimension
and $D$ the fractal dimension.
%The spanning  cluster is also called the infinite cluster, infinite incipient
%cluster etc, but we stick to the nomenclature "spanning cluster" 
%in a finite lattice. 
%However, recent studies show that there exist large clusters
%which also show the same behaviour although they do not actually span the
%lattice \cite{NSA}.  

Distribution of cluster masses at and away from criticality
has been studied in detail \cite{SA}.  
Conditional probability distributions for
spanning cluster (SC) masses, their moments,  and other variables like
the shortest path etc. also appear in the literature \cite{Hav1,Hav2,Derr,PP,HovA,NSA}.
While the distribution for the cluster masses show a power
law behaviour at criticality \cite{SA}, the  probabilities of spanning clusters
masses have an entirely different variation \cite{Hav2,HovA}. 

In this article we report a study of  the probability distribution functions 
of the masses of the spanning clusters which span the lattice from
top to bottom and  a comparative analysis for different dimensions. 
Here the condition that the
cluster spans along {\it one} particular direction of the lattice is 
necessary and
sufficient and hence the condition of the spanning along all directions
is relaxed. 

We examine the distribution functions separately for the two cases
(a) when there exists only one SC 
(b) when there are more than one  coexisting spanning clusters.
Although case (a) occurs  predominantly, case (b) has recently
been established \cite{Aizen} 
to have a finite non-zero probability of
occurrence even in two dimensions. Little is known 
about the distribution functions of masses in case (b) and 
we attempt to extract as much information for this as possible.

We have simulated $L^d$ hypercubic lattices in $d$ dimensions
with helical boundary conditions where each site is occupied 
with a probability $p$.  The clusters are identified 
 using the Hoshen Kopelman algorithm. The largest lattices 
considered  have sizes  $L = 800$ in $d = 2$, $L = 60$ in $d = 3 $, $L = 30$ in $d=4$
and $L = 15$ in $d=5$. A maximum of $10^6$ initial configurations (for the
smallest lattices) were
generated at the percolation threshold $p_c$ where  the values of $p_c$ 
given in  
ref \cite{SA} have been  used.

 As   it is known that $M$ scales as  $L^D$, 
where $D$ is the fractal dimension of the spanning cluster, 
we have directly measured the probability distribution of $M/L^D$, i.e., 
the bin sizes are chosen to be proportional to $1/L^D$. 
We normalise the probabilities so that the total probability
is unity. We 
 find that the {\it normalised } probabilities plotted against 
$m = M/L^D$ all collapse on a single curve for different system sizes. 
This happens in  all dimensions  from two to five. As an example, the collapses
 in two 
and three dimensions are shown in Fig. 1.  Finite sizes 
effects are stronger in  higher dimensions. 
%The distribution  for higher dimensions develop longer tails and become 
%narrower.  

The probability distribution  is  of  the form:
% function fits  as a   power-exp of the following form (cf. \cite{Hav2,NSA}
\begin{equation}
P(M/L^D) \propto f(M/L^D)
\end{equation}
where 
\begin{equation}
f(x)  = A x^\alpha \exp(-\gamma x^\beta)
\end{equation}

Fitting the universal scaling function in the above form is
best in two dimensions. However, for the tail of
the distribution, the above form   gives a very good fit 
even in higher dimensions. It maybe added here that 
the tail of the distribution becomes important in many problems,
e.g., in problems related to stock market fluctuations \cite{Manten}.

The form of the probability distribution
obtained here is very close to that studied in \cite{Hav2}.
We do not get any prefactor for the scaling functions 
%which is
%due to the fact that the probabilities are normalised and the
here as the bin sizes are  proportional to $1/L^D$. 
(This factor, as $1/M$ 
appears  
when the normalised probabilities are
also  divided  by the bin sizes
as in  \cite{Hav2}.) 
 However, the exponents  obtained in 
the present study  are
totally  different. 
For example, in two dimensions, $\beta = 6.7 \pm 0.1$ and $\gamma \sim 10$
while in \cite{Hav2} the corresponding values are $\sim 19$ and $\sim 10^{-8}$
respectively.
The possible reasons for this discrepancy are  discussed
later.
% which maybe due to the fact that the boundary conditions
%are entirely different. 

Quantitative comparison of the distributions for different
dimensions is done by calculating  the first and second  cumulants
of the distributions and studying their behaviour with the dimensionality. 
In each dimension, we extrapolate these results for $1/L \rightarrow 0$
 as there  
are  some finite size effects.  In general we fit  the cumulants
as linear functions of   $1/L$ to extrapolate. The extrapolated values    
vary as  simple power laws as given  below (see Fig. 2)  
\begin{equation}
\langle m\rangle \sim d^{-a}
\end{equation}
\begin{equation}
\sigma^2 = \langle m^2\rangle   - \langle m\rangle ^2 \sim d^{-2b}
\end{equation}
with $a = 1.65 \pm 0.1$
and $b = -0.25 \pm 0.01$.

We find the scaling form of the distribution by fitting with appropriate values
of $A,  \alpha,\gamma$ and $\beta$.  
Again $\alpha,  A $ and $\beta$ show simple power law variations 
with dimensionality: the powers are close to -2 for $\alpha $ and $\beta $ 
(see Fig. 2),  $\gamma$ apparently has no dependence on the
dimensionality. 

\medskip

The case when there exists more than one spanning clusters has also been
explored. 
We rank the 
spanning clusters by their sizes and  obtain separately the distributions
for the $rth$ largest cluster when the total number of spanning clusters is $n$.
For two dimensions, when there exists two SC's, 
the distribution  function for the larger SC is clearly  
different 
from that of the unique  SC (see Fig. 3). 
In particular, the distribution is more symmetric in comparison 
to that of the unique SC and more sharply peaked. 
One concludes that there is a different universal function for 
the SC's when $n > 1$. However, an attempt to find the exact form
of this distribution is difficult because of the fluctuations in
the data. 
This fluctuation is unavoidable as the probability 
of cases with $n > 1$ is very small.
 
However, certain features of the distribution  of the masses
are available from the present study. The mean value of $M/L^D$  varies 
appreciably, for example, in the
$n =2$ case in two dimensions:  for the largest SC   
$\langle m \rangle _{n=2,r=1} \sim  0.42 $
compared to 
$\langle m\rangle _{n=1,r=1} \sim   0.58$, 
where $m_{n,r}$ is the mass of the  $r$th  largest
SC when the number of SC is $n$.
Whereas, in higher dimensions,  the largest spanning clusters  in the 
multiple SC cases
become comparable in size whatever be the number of SC's. 
Such a result indicates that in the higher dimensions,
the largest SC is unaffected by the presence of others - consistent
with the fact that it is easier to conceive independent coexisting
spanning clusters along one direction in large dimensions.
%The distribution functions behave in the same way. 
                                                             
However, the width becomes smaller 
 and $\sigma$ shows a power law behaviour with $n$: 
\begin{equation}
 \sigma = (\langle m^2\rangle  - \langle m\rangle^{2})_{n,r=1}  \sim n^{-2c}
\end{equation}
with   $c \sim 0.3 \pm 0.02$.
This 
is shown for the case for five dimensions where one can obtain an appreciable
number of SC's numerically (Fig. 4).

We also get conclusive results for the ratios of the SC masses when 
e.g., $n =2$.
This  ratio is around 1.4 in case of $d=2$ (also obtained in \cite{PSAA}) 
and 2.2 for $d=5$. 
This indicates that the larger SC becomes dominant in higher dimensions.
%there is a  dominant SC in higher dimensions compared to lower.
%This result is however not unexpected, as in two dimensions, it is
%difficult to visualise a dominant spanning cluster as well as the 
%resence of others. 
%his  ratio is around 1.4 in case of 2-d (also obtained in \cite{PSAA}) and 2.2 for 5d. 

In summary, we have obtained several quantitaties related to   the distribution
of the spanning cluster masses systematically varying with the dimensionality.
A universal scaling function is obtained in each dimension, having 
similar form with dimension dependent exponents. These dependences appear
as simple power law variations of the dimensionality. 
The dependence of the exponents on the dimensionality is not surprising;
in fact, the variation of the cumulants with the dimension
is related to the dimension-dependent exponents.
One can, in principle, also fit these exponents as polynomial functions of
$(6-d)$ (as $6$ is the upper critical dimension in percolation). We keep the question of exact variation of the exponents
as function of dimension open as it is difficult  to 
obtain a concrete form from the  numerical simulations only.

As mentioned before, these
exponents do not match with some earlier results \cite{Hav2,HovA}. However, it must be noted
that in these studies, the conditional probabilities were obtained 
with different boundary conditions in the sense that the clusters
were required to span in {\it all} directions. 
There are differences in the values of $\langle m\rangle$ 
and $\sigma$ as well, $\langle m\rangle$ being smaller 
when the SC spans along one direction only.
Hence the universal function
seems to be highly dependent on the boundary conditions and in that sense
only weakly universal. We also do not attempt
to fit the scaling function by an alternative form as 
that admits yet another parameter and the fitting becomes difficult to
handle.
Another less important point is that  in  \cite {Hav2},  
the distributions  are obtained  for any number of  SC  present.
Although the distributions in the one SC case and two SC case
are quite different, it should 
however not matter as the latter has a very small probability of
occurrence.

We have also shown qualitatively that the distributions for the
SC's in multiple SC case are different from that of the one SC case.
The second  cumulant  of the largest SC, in particular, 
apparently varies as a power law with
$n$. The mean $\langle m\rangle$ for the largest SC varies appreciably
in low dimensions but becomes a constant in higher dimensions. 

The author is grateful to B. K. Chakrabarti  for  discussions
and D. Stauffer for critical comments on the manuscript. 
She  also thanks  A. Aharony for bringing ref \cite{Hav2} to notice. 
 
%\begin{references}

%{\noindent {\bf Figure captions}}

\vskip 2cm

\pagebreak
%\narrowtext
\begin{figure}

\psfig {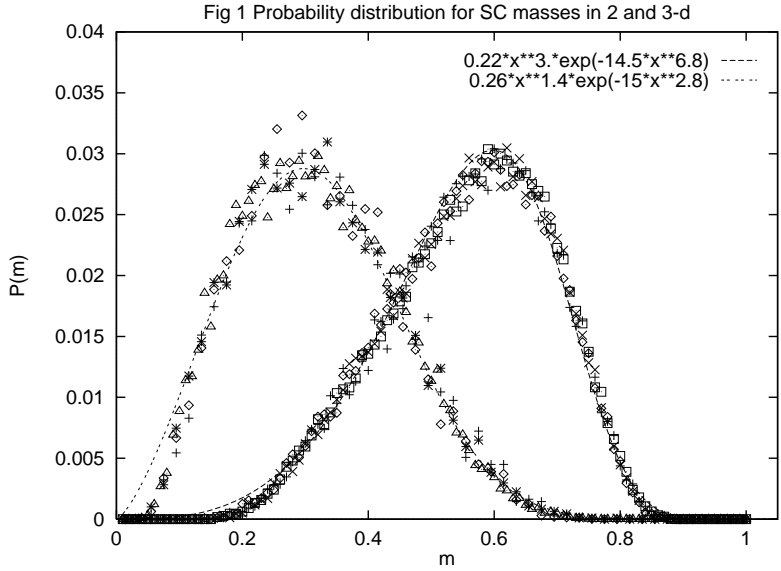}
\caption{The probability distribution for SC masses for different lattice 
sizes are shown for $d=3$ (left) and $d=2$ (right). The dashed curves are 
possible fittings for the universal functions.}
\end{figure}

\begin{figure}
\psfig {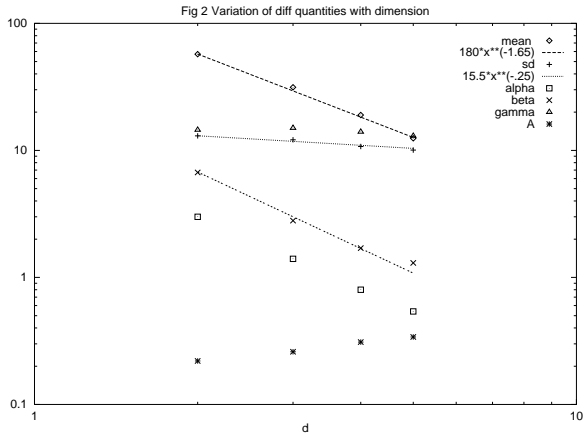}
\caption{Variations of different quantities with dimension. Power law
fittings for the first two cumulants are shown. The dashed
line has a slope equal to -2.}
\end{figure}
\begin{figure}
\psfig {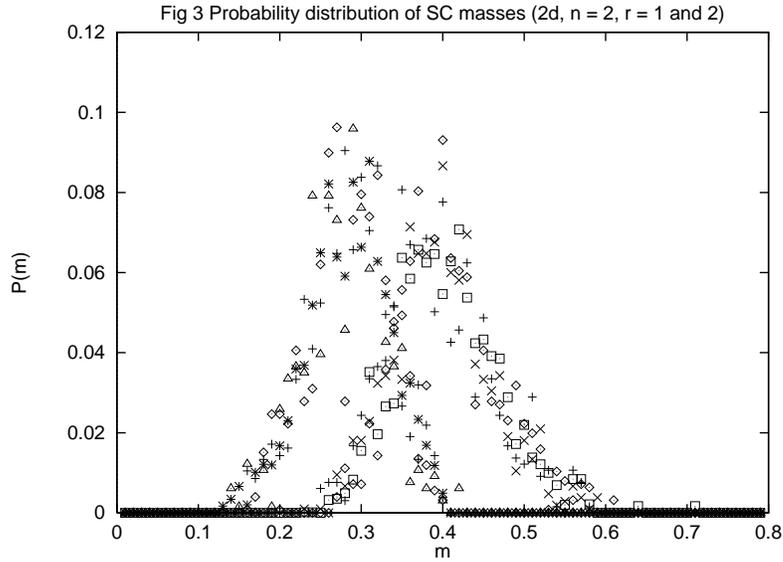}
\caption{The probability distribution for the two coexisting SC's are shown for different
lattice sizes in two dimensions. }
\end{figure}

\begin{figure}

\psfig {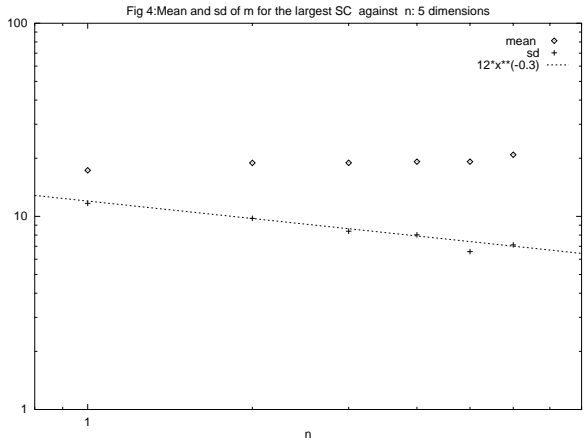}
\caption{The variation of the first two cumulants of the largest SC is shown against
the number of SC in five dimensions. The lattice size is $13^5$.}
\end{figure}

%\end{large}
\end{document}